\documentclass[aps,showpacs,superscriptadress,preprint]{revtex4}
\usepackage{graphicx}
\usepackage{amsmath}
\begin{document}
\title{Improved quark mass density- dependent model with quark and non-linear scalar field coupling}

\author{Chen Wu$^{1}$\footnote{022019003@fudan.edu.cn},
Wed-Liang Qian$^1$\footnote{wlqian@fudan.edu.cn} and Ru-Keng
Su$^{2,1,3}$\footnote{rksu@fudan.ac.cn}} \affiliation{
\small 1. Department of Physics, Fudan University, Shanghai 200433, P.R. China\\
\small 2. CCAST(World Laboratory), P.O.Box 8730, Beijing 100080, P.R. China\\
\small 3. Center of Theoretical Nuclear Physics,\\
National Laboratory of Heavy Ion Collisions, Lanzhou 730000,
P.R.China}

\begin{abstract}
The improved quark mass density- dependent model which includes
the coupling between the quarks and a non-linear scalar field is
presented. Numerical analysis of solutions of the model is
performed over a wide range of parameters. The wave functions of
ground state and the lowest one-particle excited states with even
and odd parity are given. The root-mean squared radius, the
magnetic moment and the ratio between the axial-vector and the
vector $\beta$-decay coupling constants of the nucleon are
calculated. We found that the present model is successful to
describe the properties of nucleon.
\end{abstract}

 \maketitle

\section{Introduction}
Since the conjecture of Witten [1] that strange quark matter(SQM),
would be more stable than the normal nuclear matter, much
theoretical effort has been made on the investigation of its
properties and applications [2-13]. Because of the well-known
difficulty of QCD in the non-perturbative domain, many effective
models reflecting the characteristics of the strong interaction
are  used to study the SQM. They include the MIT bag model [2-4],
the quark-meson coupling(QMC) model [5], the Friedberg-Lee(FL)
soliton bag model [6], the chiral SU(3) quark model [7], the quark
mass density-dependent(QMDD) model [8-11] and the quark mass
density- and temperature- dependent model [12, 13], etc. In this
paper, we will focus our attention on the QMDD model.

The QMDD model was first suggested by G. N. Fowler, S. Raha, and
R. M. Weiner. According to this model, the masses of $u,d$ quarks
and strange quarks (and the corresponding anti-quarks) are given
by
\begin{eqnarray}
m_{q} = \frac{B}{3n_{B}} (i = u,d,\bar u,\bar d),   \label{mq1}
\end{eqnarray}
\begin{eqnarray}
m_{s,\bar s } = m_{s0}+\frac{B}{3n_{B}},     \label{mq2}
\end{eqnarray}
where $m_{s0}$ is the current mass of the strange quark, $B$ is
the vacuum energy density inside the bag, and $n_{B} =
\frac{1}{3}(n_u+n_d+n_s)$ is the baryon number density, with
$n_u,n_d,n_s$ representing the density of $u$ quark, $d$ quark,
and $s$ quark. The basic hypothesis Eqs. (1) and (2) in QMDD model
can easily be understand from the quark confinement mechanism. A
confinement potential kr (r$^2$) must be added to a quark system
in the phenomenological effective models because  the the
perturbative QCD can not give us the confinement solutions of
quarks. The confinement potential prevents the quark from going to
infinite or to the very large regions. The large volume means that
the density is small. This mechanism of confinement can be
mimicked through the requirement that the mass of an isolated
quark becomes infinitely large so that the vacuum is unable to
support it [11, 12]. This is just the physical picture given by
Eqs (1) and (2). In fact, this confinement mechanism is very
similar to that of MIT bag model. But the advantage of QMDD model
is that it does not need to introduce a quark confined boundary
condition as that of the MIT bag model.

Although the QMDD model is  successful for describing the
properties of SQM [8-11], but it is still an ideal quark gas
model. Compared to the usual ideal quark gas model, the basic
improvement of the QMDD model is that the quark masses depend on
density and the quark confinement mechanism is mimicked.
Obviously, if we hope to investigate the physical properties of
nucleons and hyperons by means of the QMDD model, the quark-quark
interactions must be considered. Following the line of QMC model,
we will introduce  quark and scalar $\sigma$ field nonlinear
coupling self-consistently to improve the QMDD model in this
paper. As a first step, we ignore the s-quark and consider the
coupling of $u$ and $d$ quarks to a non-linear scalar field only.

We hope to emphasize there are two basic differences between our
improved quark mass density- dependent (IQMDD) model and the usual
QMC model suggested first by Guichon [14] and developed by Thomas,
saito [15] and Jennings. Jin [16]. Firstly, instead of a MIT bag
for nucleon in QMC model, we do not need a MIT bag for nucleon in
IQMDD model. The constraint of MIT bag boundary condition
disappears in our formulae because the quark confinement mechanism
has been established in Eqs. (1) and (2). This is very important
because  it should provide a reasonable starting point for
many-body calculations. Secondly, the interaction between quark
and scalar meson is limited in the bag regions for QMC model, bur
for IQMDD model, this interaction is extended to the whole free
space. In fact, our model is similar to that of the F-L model, but
instead of massless quarks in the F-L model, the masses of $u$ and
$d$ quarks in our model are given by Eqs. (1) and (2).

This paper is organized as follows. The main formulae of  IQMDD
will be given in the next section. The numerical results for the
ground state and a number of physical quantities, namely, the
root-mean-squared (rms) charge radius $r_p$, the magnetic moment
$\mu_p$, the ratio between the axial-vector and the vector
$\beta$-decay coupling constants of the nucleon $g_A/g_V$ will be
presented in the third section. In the fourth section we will
study the lowest one particle excited states with even or odd
parity. The last section is a summary and discussion.

\section{The Improved quark mass density- dependent Model}
We now briefly outline the model and its pertinent features below.
The Hamiltonian density of the IQMDD model reads:
\begin{eqnarray}
H ={\psi ^+}
[\frac{1}{i}{\overrightarrow{\alpha}}{\cdot}{\overrightarrow{\nabla}}
+\beta(m_{q}+f\sigma)]\psi+\frac{1}{2}\pi^2+\frac{1}{2}(\nabla\sigma)^{2}+U(\sigma),
\end{eqnarray}
where $m_q=B/{3n_B}$ is the mass of u(d) quark,
$\overrightarrow{\alpha}$ and $\beta$ are the standard Dirac
matrices , $\psi$ represents the quark quantum field (color and
flavor indices suppressed) satisfying the canonical
anticommutation relations:
\begin{eqnarray}
\{\bar \psi(\overrightarrow r,t),\psi ({\overrightarrow r}^{'},t)
\}=\delta^3(\overrightarrow{r}-\overrightarrow{r}^{'}),
\end{eqnarray}
and $f$ is the coupling constant between the quark field $\psi$
and the meson field $\sigma$. The $\sigma$ field is considered
independent of time and consequently the commutator:
\begin{eqnarray}
[{\pi} (\overrightarrow r),\sigma ({\overrightarrow r}^{'})]=0,
\end{eqnarray}
where $\pi$ is conjugate field of the scalar meson field. Hence,
$\sigma$ is treated as a classical field. In Eq. (3) $U(\sigma)$
is the self interaction potential for $\sigma$ field, which has a
phenomenological form:
\begin{eqnarray}
U(\sigma)=\frac{c_2}{2}{\sigma}^2+\frac{c_3}{6}{\sigma}^3+\frac{c_4}{24}{\sigma}^4+B,
\end{eqnarray}
and
\begin{eqnarray}
c_3^{2} > 3c_2c_4,
\end{eqnarray}
to ensure that the absolute minimum of $U(\sigma)$ is at
$\sigma=\sigma_{vac}\neq 0$. The bag constant $B$ is introduced in
order that
\begin{eqnarray}
U(\sigma_{vac})=0\ \ {\ and}\ \ U(0)=B.
\end{eqnarray}

Without any lose of generality, we may choose $c_3<0$, and
therefore $\sigma_{vac} > 0$:
\begin{eqnarray}
\sigma_{vac}=\frac{3}{2c_4}[-c_3+({c_3}^2-\frac{8}{3}{c_2}{c_4})^{1/2}].
\end{eqnarray}

One can construct a Fock space of quark states and expand the
operator $\psi$ in terms of annihilation and creation operators on
this space with c-number spinor functions $\varphi_{n}^{(\pm)}$,
which satisfies the Dirac equation:
\begin{eqnarray}
[\frac{1}{i}{\overrightarrow{\alpha}}{\cdot}{\overrightarrow{\nabla}}
+\beta(m_{q}+f\sigma)]\varphi_{n}^{(\pm)}=\pm\epsilon_{i}\varphi_{n}^{(\pm)}
\label{e1}
\end{eqnarray}
with superscripts ${\pm}$ denoting the positive and negative
energy solutions, respectively. The spinor functions $\varphi_n$
are normalized according to
\begin{eqnarray}
\int {\varphi}_n^+ {\varphi}_n d^3r=1.
\end{eqnarray}

The total energy  of the quark-scalar field system is given by
\begin{eqnarray}
E(\sigma)=\sum_n\epsilon_n
+\int[\frac{1}{2}(\nabla\sigma)^{2}+U(\sigma)]{d^{3}}r.
\end{eqnarray}
Minimum of $E(\sigma)$ occurs when $\sigma$ is the solution of
\begin{eqnarray}
-\nabla^{2}\sigma+\frac{dU(\sigma)}{d\sigma}=-f\sum_n\bar{\varphi}_n\varphi_n.
\label{e2}
\end{eqnarray}

The equations of the quark field for ground state and excited
states will be given in Sec. 3 and Sec. 4 respectively.

\section{The Ground State Solution}
We discuss the ground state solution of the system now. Define
$\varphi$ as the lowest positive energy wave function. It can be
expressed as:
\begin{eqnarray}
\varphi=\left(\begin{array}{c} u \\
i(\frac{\overrightarrow{\sigma}\cdot\overrightarrow{r}}{r})v
\end{array}\right)\chi_{m},                             \label{wf}
\end{eqnarray}
where $\overrightarrow\sigma$ are the Pauli matrices,
$\chi_{m}=\left(\begin{array}{c}1
\\0\end{array}\right)$ or $\left(\begin{array}{c}0\\1\end{array}\right)$.
The radial functions $u(r)$ and $v(r)$ satisfy:
\begin{equation}
\frac{du(r)}{dr}=-[{\epsilon}+{m_q}+f{\sigma}(r)]v(r),
\end{equation}
\begin{equation}
\frac{dv(r)}{dr}=-2\frac{v(r)}{r}+[{\epsilon}-{m_q}-f{\sigma}(r)]u(r),
\end{equation}
the normalized condition reads:
\begin{equation}
4\pi\int_0^{\infty}[{u}^2(r)+{v}^2(r)]r^2dr=1,
\end{equation}
and the equation of motion of $\sigma$ field Eq. (13) becomes:
\begin{equation}
\frac{d^2\sigma(r)}{dr^2}+\frac{2}{r}\frac{\sigma(r)}{dr}
=U^{\prime}(\sigma(r))+3f[{u}^2(r)-{v}^2(r)],
\end{equation}
Equations (15), (16) and (18) can be solved with the boundary
conditions:
\begin{equation}
v(r=0)=0,\ \ u(r=\infty)=0,\ \ v(r=\infty)=0,
\\ \ \ \sigma(r=\infty)=\sigma_{V}, \ \ \sigma^{\prime}(r=0)=0
\end{equation}
 Before numerical calculation, we address the parameters of the
model first. There are four free parameters, namely, $c_2$, $c_3$,
$c_4$, and $f$ in IQMDD model. The parameters $c_2$, $c_3$ and
$c_4$ fix the interaction potential $U$, while $f$ measures the
coupling between the quark and the scalar field. The set of
equations (15), (16) and (18) should be solved alternately until
consistency was obtained using the iterative method [17]. Once the
solutions of the above equations are obtained, one can calculate a
number of physical quantities pertaining to the three-quark system
which have been measured from experiments. Let $r_{p}$, $\mu_{p}$,
and $g_{A}/g_{V}$ be, respectively, the rms charge radius, the
magnetic moment, and the ratio between the axial-vector and the
vector $\beta$-decay coupling constants of the nucleon. They
satisfy [18]:
\begin{eqnarray}
<{r_{p}}^{2}>=\int \varphi^{+}\varphi{r}^{2}d^{3}r/\int
\varphi^{+}\varphi{d^{3}}r,
\end{eqnarray}
\begin{eqnarray}
{\mu_{p}}=\frac{1}{2}(\int
\vec{r}\times\varphi^{+}\vec{\alpha}\varphi{d^{3}r})_{z}/\int
\varphi^{+}\varphi{d^{3}}r,
\end{eqnarray}
and
\begin{eqnarray}
g_{A}/g_{V}=\frac{5}{3}\int
\varphi^{+}\sigma_{z}\varphi{d^{3}r}/\int
\varphi^{+}\varphi{d^{3}r},
\end{eqnarray}
By using Eq. (14), we find:
\begin{eqnarray}
<{r_{p}}^{2}>=4\pi\int_0^{\infty}(u^{2}+v^{2})r^{4}dr,
\end{eqnarray}
\begin{eqnarray}
\mu_{p}=\frac{8\pi}{3}\int_0^{\infty} r^{3}uvdr,
\end{eqnarray}
and
\begin{eqnarray}
g_{A}/g_{V}=\frac{20\pi}{3}\int_0^{\infty}r^{2}(u^{2}-\frac{1}{3}v^{2})dr.
\end{eqnarray}

Since in the present form the model is flavor independent, the
charge radius of the neutron in the model is
\begin{eqnarray}
<r_n^2>\equiv 0 ,
\end{eqnarray}
and the neutron magnetic moment is
\begin{eqnarray}
{\mu}_n=-\frac{2}{3}{\mu}_p ,
\end{eqnarray}
as given by the $SU(6)$ algebra. Corrections to these relations
arise only when QCD effects are included.

For comparison let us remind ourselves that the experimental
values of proton mass $E = 4.69$ fm$^{-1}$, the proton magnetic
moment $\mu = 2.79$ nuclear magnetons, $g_{A}/g_V = 1.25$, and the
charge radius of proton $r_{p} = 0.83$ fm. To do numerical
calculation, we follow Ref. [17, 18]. We fix $ r_{p} = 0.83$ fm
first. Once the value is chosen, there are only three free
parameters. We have studied mostly two families of parameters for
ground system state. These are the $c_2 = 0$ and the $B^{1/4} =
145$ MeV cases respectively. Each choice fixes one other parameter
by a specific requirement for the shape of the potential.
Therefore it is sufficient to label our results by the coupling
constant $f$ and constant $c_{4}$.

Our results for the two families $(c_2 = 0, B^{1/4} = 145$ MeV)
are summarized in three tables. Through these tables, we use the
units $\hbar = c = 1$ and use fm as the fundamental unit for
length.

We consider the case $B^{1/4}=145$ MeV first. The bag constant $B$
corresponds to the difference between two minima of the
$U(\sigma)$. In Table. 1 we list the bag properties as a function
of the parameter $c_4$ for two values of the coupling constant
$f$. The variation of bag properties with the coupling constant
$f$ for several values of $c_4$ is given in Table. 2. Several
features emerge from these calculations. Firstly, we note that an
increase of the coupling constant produces a continuous change
from a volume quark distribution for small $f$ to a surface quark
distribution for large $f$. This change is illustrated in Fig. 1
where we plot the quark density $u^2-v^2$ for three values of $f$.
It should be noted that shape of the soliton field does not change
significantly with $f$ for given $c_4$. The variations of the
quark charge density as a function of the radius with increasing
$c_4$ are plotted in Fig. 2 and Fig. 3. When $c_4$ decreases, with
a fixed value of $f$, the quark charge distribution $u^2-v^2$
change slowly from surface to volume. This is illustrated in Fig.
2 and Fig. 3 for $f = 75$ and $f = 200$, respectively. The change
from volume to surface quark charge density is also evident in the
variations of the values of the magnetic moment $\mu_p$ and
$g_A/g_V$. For instance from Table. 2, when $c_4$ = $8\times
10^5$, the magnetic moment varies with increasing $f$ from
$1.87\mu_B$ to $2.61\mu_B$, where $\mu_B$ is the Bohr magneton of
the proton. Similarly, $g_A/g_V$ varies from 1.21 for small $f$ to
0.618 for large $f$. It is also found from the tables that for
some $c_4$ and $f$ cases we can not found a solution for $r_p =
0.83$ fm.

The variation of  the scalar field $\sigma$ as a function of the
radius for $f=50$, $c_4=2\times 10^5$ and $B^{1/4}=145$ MeV is
presented in Fig. 4. The value of $\sigma$ inside the hadron is
very different from that of outside: inside $\sigma$ is less than
zero, but outside $\sigma$ approaches $\sigma_{vac}$. The abrupt
transition of scalar field through hadron surface will contribute
to the total energy remarkably, as shown in Eqs. (11) and (12).
 \newline\\
\begin{tabular}
{p{1.7cm}p{2.0cm}p{2.0cm}p{2.0cm}p{2.0cm}p{1.0cm}}
\multicolumn{6}{c}{TABLE 1.  Variation of the properties with
increasing parameters  } \\
\multicolumn{6}{c}{
 for the $B^{1/4}=145$ MeV bag for the two values of $f$.} \\
\hline\hline $f$ &$c_4$ &${\epsilon}$  &$E$
&$\mu_{p}$&$g_{A}/g_{V}$
\\\hline
30 &$8\times10^4$  &1.54  &6.54  &2.39 &0.904  \\
              &$1\times10^5$  &1.56  &6.56   &2.38 &0.929 \\
              &$4\times10^5$  &1.74  &6.94  &2.25 &1.06 \\
              &$8\times10^5$  &1.85  &7.26  &2.15 &1.11 \\

200 &$8\times10^4$  &\multicolumn{4}{c}{No solution}  \\
              &$1\times10^5$  &\multicolumn{4}{c}{No solution} \\
              &$4\times10^5$  &1.24  &5.23  &2.64 &0.603 \\
              &$8\times10^5$  &1.22  &5.21  &2.63 &0.618 \\
 \hline\hline
\end{tabular}

\begin{tabular}
{p{1.7cm}p{2.0cm}p{2.0cm}p{2.0cm}p{2.0cm}p{1.0cm}}
\multicolumn{6}{c}{\ \  TABLE 2.  Variation of the properties as a
function of $f$ for  } \\
\multicolumn{6}{l}{
several values of $c_4$ with $B^{1/4}=145$ MeV.} \\
\hline\hline $c_4$ &$f$ &${\epsilon}$  &$E$
&$\mu_{p}$&$g_{A}/g_{V}$
\\
\hline
$8\times10^5$ &15  &2.18  &8.16  &1.87 &1.21  \\
              &30  &1.85  &7.26   &2.14 &1.11 \\
              &50  &1.56  &6.47  &2.39 &0.936 \\
              &75  &1.41  &5.97  &2.49 &0.794 \\
              &100  &1.24  &5.70  &2.55 &0.720 \\
              &200  &1.22  &5.21  &2.61 &0.618 \\

$4\times10^5$ &15  &2.11  &8.01  &1.96 &1.22  \\
              &30  &1.74  &6.94  &2.25 &1.06 \\
              &50  &1.49  &6.28  &2.43 &0.868 \\
              &75  &1.36  &6.07 &2.55 &0.743 \\
              &100  &1.30  &5.81  &2.57 &0.680 \\
              &200  &1.24  &5.23  &2.63 &0.603 \\

$1\times10^5$ &15  &1.96  &7.62  &2.08 &1.17  \\
              &30  &1.56  &6.56  &2.39 &0.929 \\
              &50  &1.41  &6.09  &2.48 &0.765 \\
              &75  &1.33  &5.77  &2.50 &0.677 \\
              &100  &\multicolumn{4}{c}{No solution}\\
              &200  &\multicolumn{4}{c}{No solution}\\

 $8\times10^4$ &15  &1.94  &7.61  &2.09 &1.16  \\
              &30  &1.54  &6.54  &2.39 &0.904 \\
              &50  &\multicolumn{4}{c}{No solution} \\
              &75   &\multicolumn{4}{c}{No solution}\\
              &100   &\multicolumn{4}{c}{No solution}\\
              &200   &\multicolumn{4}{c}{No solution}\\
 \hline\hline
\end{tabular}
\\\\

Another family of parameters considered is characterized by $c_2 =
0$. In this case U($\sigma$) has an inflection point at
$\sigma=0$, and only one minimum. We vary $c_3$, $c_4$ and $f$
subject to the bag size constraint $r_p = 0.83$ fm. Our numerical
results for this family are summarized in Table. 3. We  find that
when  $f$ increases, the character of the bag changes from volume
confinement to surface confinement again. Similarly, we also find
that for a given $f$, when $c_4$ increases, the magnetic moment
decreases. In summary, we find in both cases:

(1)The quark energy $\epsilon$ and the ratio $g_A/g_V$ decreases
with increasing parameters $f$ and $c_4$.

(2)On the other hand, the magnetic moment $\mu$, has just the
opposite behavior: it is a decreasing function of the parameter
$c_4$ and an increasing function of the parameter $f$.

(3)The total energy of the ground state decreases as a function of
$f$ and increases as a function of $c_4$. The rise of the total
energy $E$ is caused by a contribution from $3\epsilon$ mainly.

(4)An increase of the coupling constant produces a continuous
change from a volume quark distribution to a surface quark
distribution, and an decrease of $c_4$ can reproduce same results.
  \newline

\begin{tabular}
{p{1.7cm}p{2.0cm}p{2.0cm}p{2.0cm}p{2.0cm}p{1.0cm}}
\multicolumn{6}{c}{TABLE 3.  Variation of the properties as a
function of $f$ and $c_4$ with $c_2$=0.} \\

\hline\hline $c_4$ &$f$ &${\epsilon}$  &$E$
&$\mu_{p}$&$g_{A}/g_{V}$
\\\hline

$1\times10^4$ &15  &1.64  &7.00  &2.31 &1.00  \\
              &30  &1.38  &6.17  &2.50 &0.769 \\
              &50  &1.29  &5.77  &2.57 &0.667 \\
              &75  &1.25  &5.53  &2.60 &0.621 \\
              &100  &1.24  &5.43  &2.61 &0.601 \\
              &150  &1.22  &5.31  &2.63 &0.583 \\

$2\times10^4$ &15  &1.64  &7.00  &2.23 &1.05  \\
              &30  &1.38  &6.17  &2.48 &0.810 \\
              &50  &1.29  &5.77  &2.57 &0.692 \\
              &75  &1.25  &5.53  &2.60 &0.635 \\
              &100  &1.24  &5.43  &2.63 &0.609 \\
              &150  &1.22  &5.31  &2.63 &0.587 \\

$4\times10^4$ &15  &1.83  &7.36  &2.16 &1.12  \\
              &30  &1.48  &6.41  &2.46 &0.857 \\
              &50  &1.34  &5.97  &2.55 &0.723 \\
              &75  &1.28  &5.71  &2.60 &0.652 \\
              &100  &1.25  &5.57  &2.63 &0.620 \\
              &150  &1.24  &5.51  &2.61 &0.592 \\
 \hline\hline
\end{tabular}

\section{The One Particle Excited States}
The one particle excited states for F-L model had been calculated
by Saly and Sundaresan [19]. In this section, we use their method
to study the one particle excited state for IQMDD model.
\subsection{First excited state with even parity}
In this configuration, we shall consider two quarks to be in the
ground state, while one quark will be placed in the first excited
level $\epsilon_1$. Then the system of equations to be solved is

\setcounter{equation}{25}
\[
\frac{du(r)}{dr}=-[{\epsilon_0}+{m_q}+f{\sigma}(r)]v(r),
\eqno(28.a)\]
\[\frac{dv(r)}{dr}=-2\frac{v(r)}{r}+[{\epsilon_0}-{m_q}-f{\sigma}(r))]u(r), \eqno(28.b)\]
\[ \frac{du_1(r)}{dr}=-[{\epsilon_1}+{m_q}+f{\sigma}(r)]v_1(r), \eqno(28.c) \]
\[\frac{dv_1(r)}{dr}=-2\frac{v_1(r)}{r}+[{\epsilon_1}-{m_q}-f{\sigma}(r)]u_1(r),
\eqno(28.d)
\]
\[
4\pi\int_0^{\infty}[u^2(r)+v^2(r)]r^2dr=1, \eqno(28.e)\]
\[4\pi\int_0^{\infty}[u^2_1(r)+v^2_1(r)]r^2dr=1, \eqno(28.f)
\]
\[
\frac{d^2\sigma(r)}{dr^2}+\frac{2}{r}\frac{d\sigma(r)}{dr}
=U^{\prime}(\sigma)+2f[u^2(r)-v^2(r)]+f[u_1^2(r)-v_1^2(r)].
\eqno(28.g)
\]

The set of equations (28.a)-(28.g) is  to be solved with the
boundary conditions: \setcounter{equation}{28}
\begin{equation}\begin{split}
 &v(r=0)=0, v_1(r=0)=0, u(r=\infty)=0, v(r=\infty)=0,\\
 &v_1(r=\infty)=0, u_1(r=\infty)=0,\sigma(r=\infty)=\sigma_{V},\sigma^{\prime}(r=0)=0.
\\
\end{split}\end{equation}
\subsection{Lowest energy with odd-parity state }
To obtain the lowest energy odd-parity state, we place two quarks
in the ground state of even-parity and one quark in the lowest
odd-parity state. In this case the system of equation is:

\[
\frac{d\widetilde{u}(r)}{dr
}=-[{\widetilde{\epsilon_0}}+{m_q}+f{\sigma}(r)]\widetilde{v}(r),\eqno(30.a)
\label{ex_odd_1}
\]\[
\frac{d\widetilde{v}(r)}{dr}=-2\frac{\widetilde{v}(r)}{r}
+[\widetilde{{\epsilon_0}}-{m_q}-f{\sigma}(r)]\widetilde{u}(r),\eqno(30.b)\]
\[\frac{d\widetilde{v_1}(r)}{dr}
=[\widetilde{\epsilon_1}-{m_q}-f{\sigma}(r)]\widetilde{u_1}(r),\eqno(30.c)
\]\[
\frac{d\widetilde{u_1}(r)}{dr}=-2\frac{\widetilde{u_1}(r)}{r}
-[\widetilde{\epsilon_1}+{m_q}+f{\sigma}(r)]\widetilde{v_1}(r),\eqno(30.d)\]
\[
4\pi\int_0^{\infty}[\widetilde{u}^2(r)+\widetilde{v}^2(r)]r^2dr=1,\eqno(30.e)\]
\[4\pi\int_0^{\infty}[\widetilde{u_1}^2(r)+\widetilde{v_1}^2(r)]r^2dr=1,\eqno(30.f)
\]
\[
\frac{d^2\sigma(r)}{dr^2}+\frac{2}{r}\frac{d\sigma(r)}{dr}=U^{'}(\sigma)+2f
[\widetilde{u}^2(r)-\widetilde{v}^2(r)]+f[\widetilde{u_1}^2(r)-\widetilde{v_1}^2(r)],\eqno(30.g)
\]
where $\widetilde{u_1}(r)$, $\widetilde{v_1}(r)$ are the
components of wave function $\varphi$:\setcounter{equation}{30}
 \begin{eqnarray}
\varphi=\left(\begin{array}{c}
\frac{\overrightarrow{\sigma}\cdot\overrightarrow{r}}{r}\widetilde{u_1}(r) \\
0\\
i\widetilde{v_1}(r)\\
0
\end{array}\right).
\end{eqnarray}

The set of equations (30.a)-(30.g) is to be solved with the
boundary conditions:
\begin{equation}\begin{split}
 &\widetilde{v}(r=0)=0, \widetilde{u_1}(r=0)=0, \widetilde{u}(r=\infty)=0,
 \widetilde{v}(r=\infty)=0,\\
 &\widetilde{u_1}(r=\infty)=0, \widetilde{v_1}(r=\infty)=0,
 \sigma(r=\infty)=\sigma_{V},\sigma^{\prime}(r=0)=0.
\\
\end{split}\end{equation}

For simplicity, we consider $ B^{1/4} = 145$ MeV, $r_p$ = 0.83 fm
case only.

\begin{tabular}
{p{1.6cm}p{1.6cm}p{1.6cm}p{1.6cm}p{1.6cm}p{1.6cm}p{1.6cm}p{1.6cm}p{1.6cm}}
\multicolumn{9}{c}{TABLE 4. Dependence of the solutions on
  $c_4$ for $f$=30.} \\
\hline\hline
 \multicolumn{3}{c}{Ground state}
 &\multicolumn{3}{c}{Even-parity state}
  &\multicolumn{3}{c}{odd-parity state}\\
$c_4$ &${\epsilon}$ &E &${\epsilon_0}$ &${\epsilon_1}$ &E$_+$
&$\widetilde{{\epsilon_0}}$ &$\widetilde{{\epsilon_1}}$ &E$_-$\\
\hline
$8\times10^5$  &1.85  &7.26  &2.02 &3.92 &9.11 &1.87 &3.10 &8.70 \\
$6\times10^5$  &1.81  &7.16   &1.99 &3.99 &9.17 &1.84 &3.11 &8.71\\
$4\times10^5$  &1.76  &7.09   &1.95 &4.12 &9.29  &1.80 &3.11 &8.74 \\
$2\times10^5$  &1.63  &6.72   &1.82 &4.16 &9.13 &\multicolumn{3}{c}{No solution}\\
$1\times10^5$  &1.56  &6.56  &1.72 &4.43 &9.36 &\multicolumn{3}{c}{No solution} \\
$8\times10^4$  &1.54  &6.54  &1.58 &4.41 &9.84 &\multicolumn{3}{c}{No solution}\\
\hline\hline
\end{tabular}

\begin{tabular}
{p{1.6cm}p{1.6cm}p{1.6cm}p{1.6cm}p{1.6cm}p{1.6cm}p{1.6cm}p{1.6cm}p{1.6cm}}
\multicolumn{9}{c}{TABLE 5.  Dependence of the solutions on
 $f$ for $c_4$=$4\times10^5$.} \\
\hline\hline
 \multicolumn{3}{c}{Ground state}
 &\multicolumn{3}{c}{Even-parity state}
  &\multicolumn{3}{c}{odd-parity state}\\
$f$ &${\epsilon}$ &E &${\epsilon_0}$ &${\epsilon_1}$ &E$_+$
&$\widetilde{{\epsilon_0}}$ &$\widetilde{{\epsilon_1}}$ &E$_-$\\
\hline
15  &2.09  &7.93  &\multicolumn{3}{c}{No solution} &\multicolumn{3}{c}{No solution} \\
20  &1.95  &7.53   &2.18  &3.80  &9.17      &2.01 &3.21 &8.82\\
25  &1.85  &7.32   &2.05 &3.91 &9.21  &1.88 &3.14 &8.79 \\
30  &1.76  &7.09   &1.95 &4.12 &9.29 &\multicolumn{3}{c}{No solution}\\
35  &1.66  &6.79  &1.80 &4.32 &9.29 &\multicolumn{3}{c}{No solution} \\
40  &1.59  &6.59  &1.59 &4.28 &9.60 &\multicolumn{3}{c}{No solution}\\
50  &1.49  &6.30  &1.48 &4.19 &9.63 &\multicolumn{3}{c}{No solution}\\
\hline\hline
\end{tabular}
\\\newline

Our results are summarized in Tables. 4, 5 and Figs. 5, 6. In
Table. 4,we fix $r_p=0.83$ fm, $B^{1/4}$=145 MeV, $f$ = 30 and
show the dependence of ground state energy E, the first
even-parity excited state energy E$_+$, and the first odd-parity
state energy
 E$_-$ on $c_4$. The dependence of E, E$_+$, E$_-$ on $f$ for $c_4$
= $4\times10^5$ is shown in Table. 5. Remember that the
experimental values of (uud) system are E = 4.69 fm$^{-1}$, E$_+$
= 7.35 fm$^{-1}$ and E$_-$ = 7.67 fm$^{-1}$ and compare our
results with that given by Table. VI, VII of Ref. [19] for F-L
model with massless quarks, we find that the agreement with
experiments for IQMDD model is better than that of F-L model.

Finally, we hope to point, as shown in Tables. 4 and 5, that the
allowed range of parameters is large for the existence of the
first even-parity excited state. The only restriction is that
 the height of the potential well $f\sigma_V$ must be
greater than the quark eigenvalues $\epsilon_1$. But for
odd-parity states, the allowed range of parameters restricts
severely. We cannot find solution in many cases. To show this
point more transparently, for odd-parity solutions, we fix
$r_p=0.83$ fm, $B^{1/4}$=145 MeV, $f$ = 30 and draw the curves of
total quark density $[2(u^2-v^2)+\tilde{u_1}^2-\tilde{v_1}^2]$ vs.
r and scalar field $\sigma$ vs. r for different $c_4$ =
$4\times10^5$, $6\times10^5$ and $8\times10^5$ in Fig. 5
respectively. We find no solutions exist for smaller values of
$c_4$. The same curves for fix $c_4$ = $4\times10^5$ but different
$f$ = 20 and 25 are shown in Fig. 6. We also find no solutions
exist for higher values of $f$.

\section{Summary and discussion}
After introducing the quark and non-linear scalar field coupling,
we suggest an improved quark mass density- dependent model. We
obtain soliton solutions of ground state and excited states  for
the coupled equations for quark and scalar fields satisfying the
required boundary conditions by numerical method. Since the most
interesting question is whether the results of these calculations
resemble the physics we are trying to describe, we concentrate on
the dependence of the results on the phenomenological parameters
introduced in this model. We present these dependence in the form
of tables and figures for the ground state and low-lying excited
states. The wave functions of quark are given. By using the wave
function of ground state, we have calculated the rms charge
radius, the magnetic moment and the ratio between the axial-vector
and the vector $\beta$-decay coupling constant of the nucleon and
compared these values with experiment. We find the results given
by IQMDD model are in agreement with experiment.

Since the boundary condition of MIT bag model has been given up in
IQMDD model, the many-body calculation beyond mean field
approximation can easily be carried out in this model.

We note that the study of this paper is still limited at zero
temperature and u, d quarks. Since the spontaneous breaking
symmetry of $U(\sigma)$ will be restored at finite temperature, it
is of interest to extend our discussion to finite temperature and
study the effect of s quarks for hyperons. Working on this topic
is in progress.

\section*{Acknowledgments}
This work is supported in part by National Natural Science
Foundation of China under No.10375013, 10347107, 10405008,
10247001, 10235030, National Basic Research Program of China
2003CB716300, and the Foundation of Education Ministry of China
20030246005.

\begin{figure}[tbp]
\includegraphics[width=10cm,height=15cm]{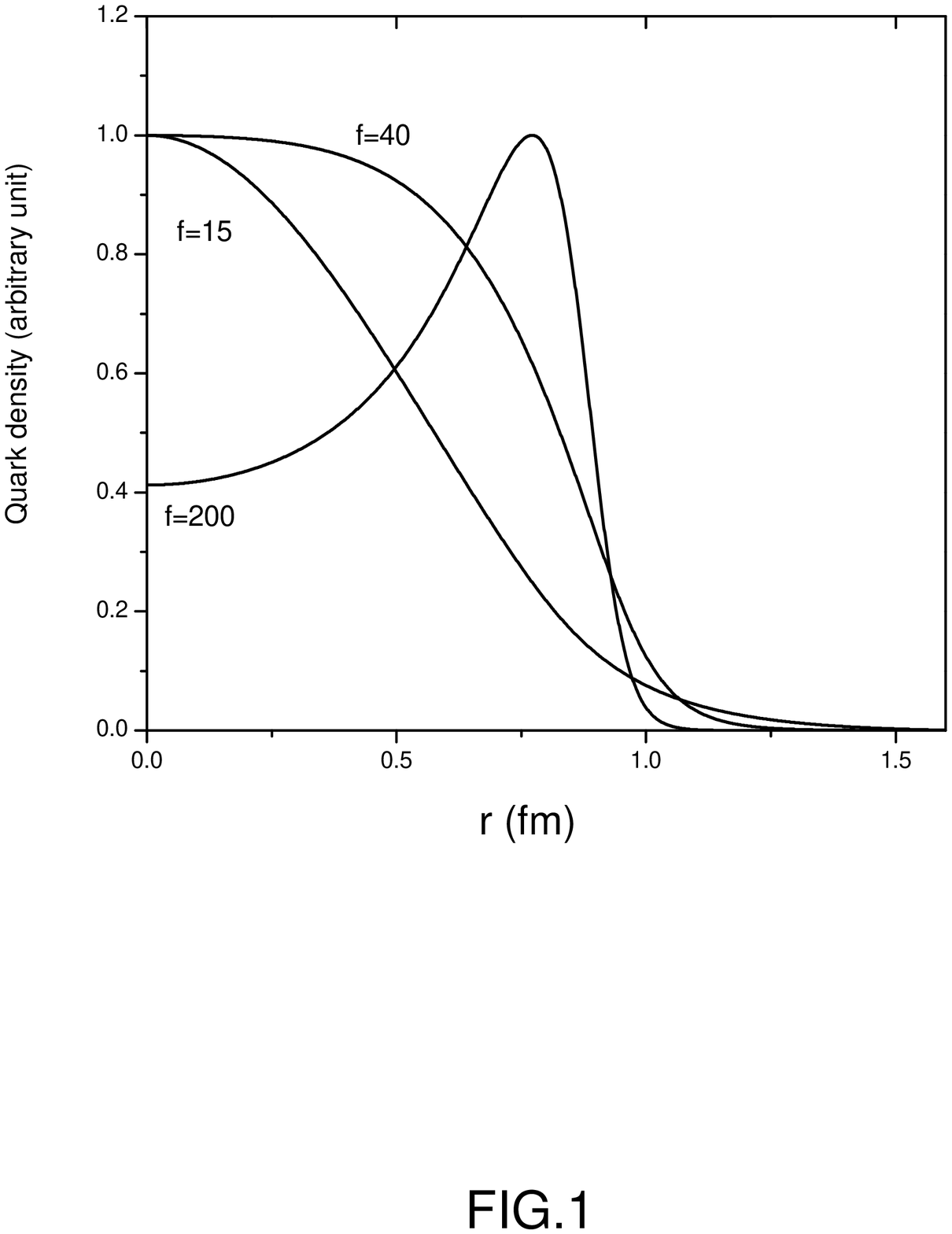}
\caption{quark density $u^2-v^2$ versus radius for $f$=15, 40, 200
 for $B^{1/4}=145$ MeV.}
 \label{fig1}
\end{figure}

\begin{figure}[tbp]
\includegraphics[width=10cm,height=15cm]{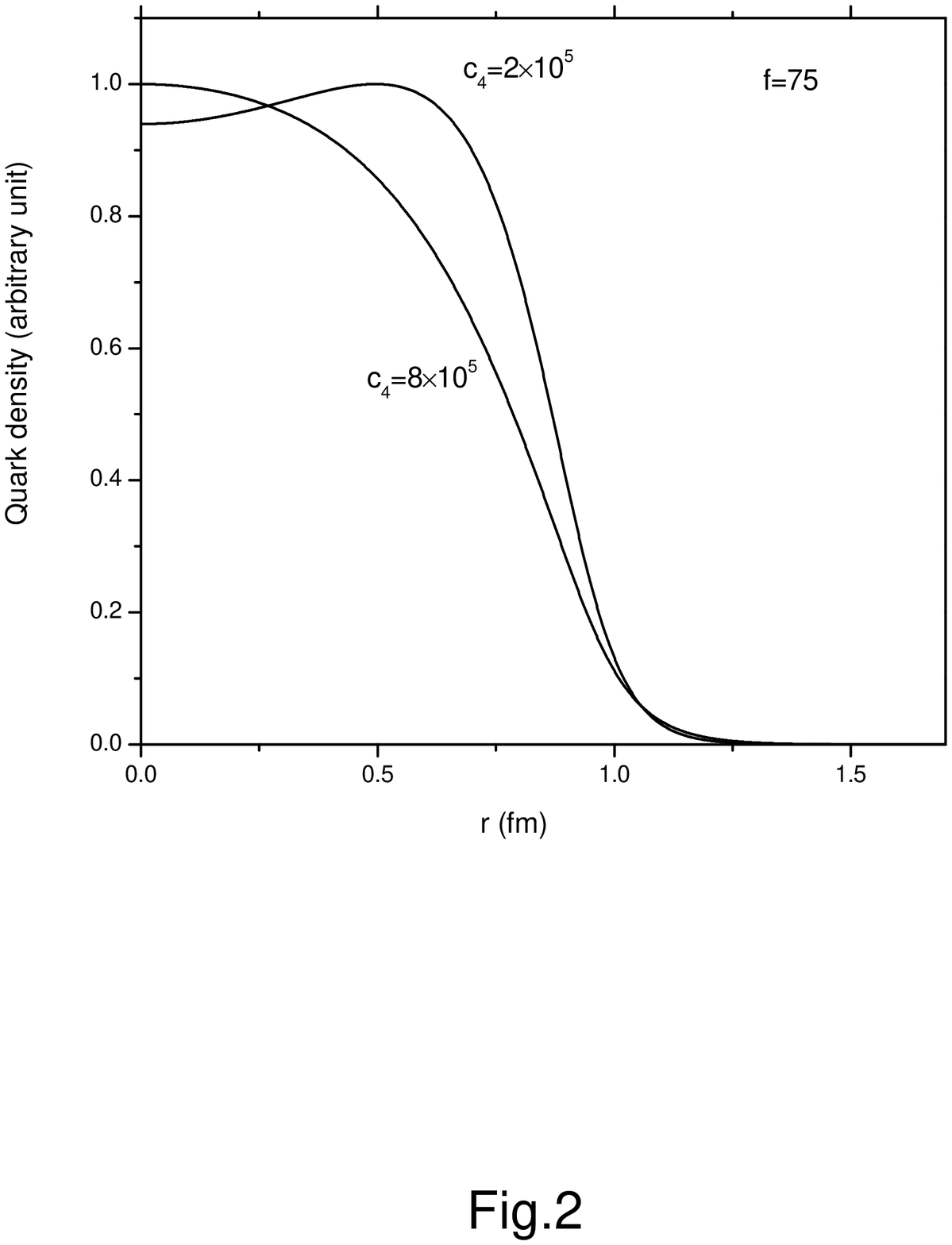}
\caption{Quark density $u^2-v^2$  versus  radius for $c_4=2\times
10^5$ and $8\times 10^5$, $B^{1/4}$=145 MeV and $f$=75.  }
\label{fig2}
\end{figure}

\begin{figure}[tbp]
\includegraphics[width=10cm,height=13cm]{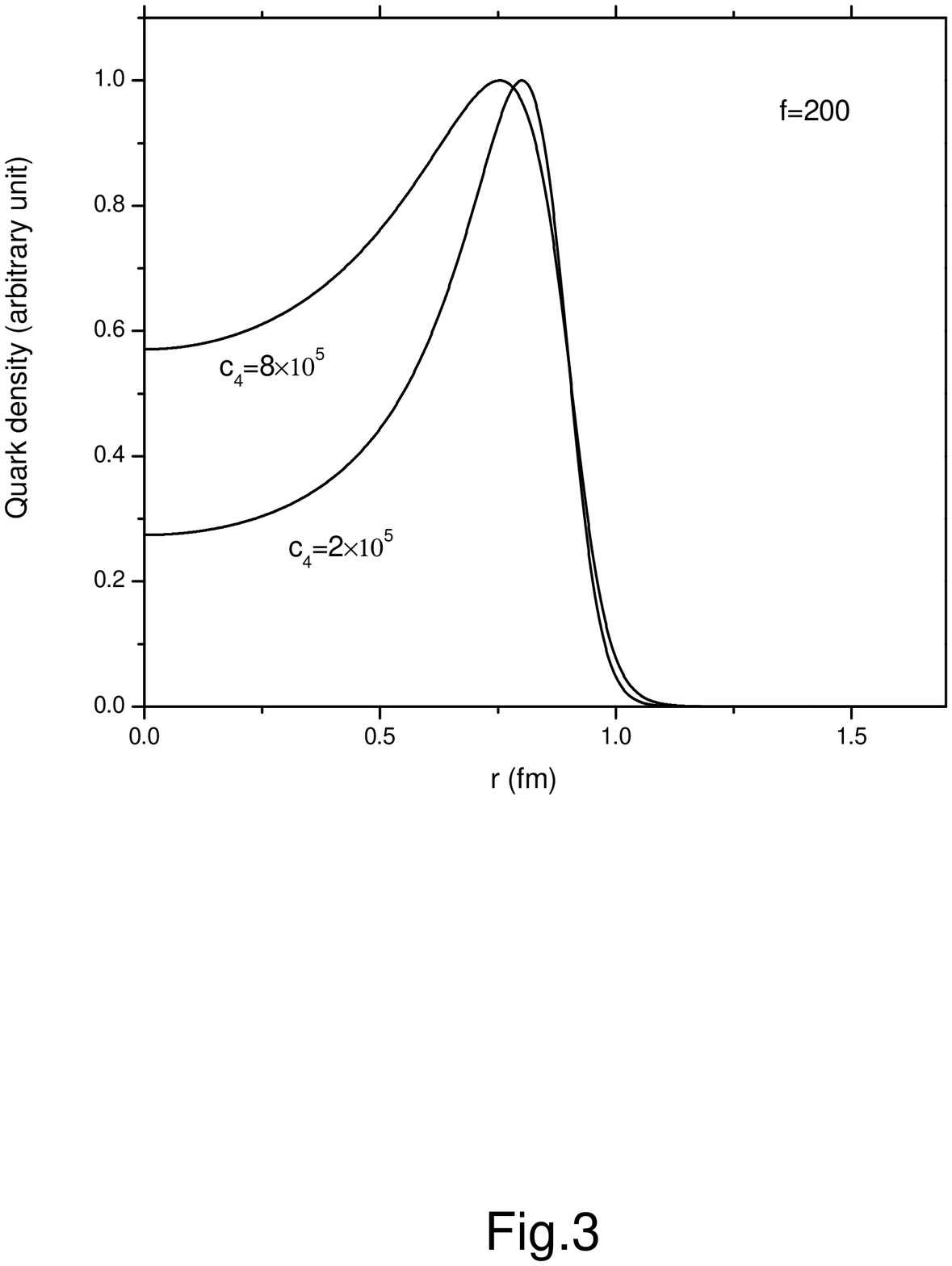}
\caption{The same as Fig.2, but $f$=200.}
 \label{fig3}
\end{figure}

\begin{figure}[tbp]
\includegraphics[width=10cm,height=15cm]{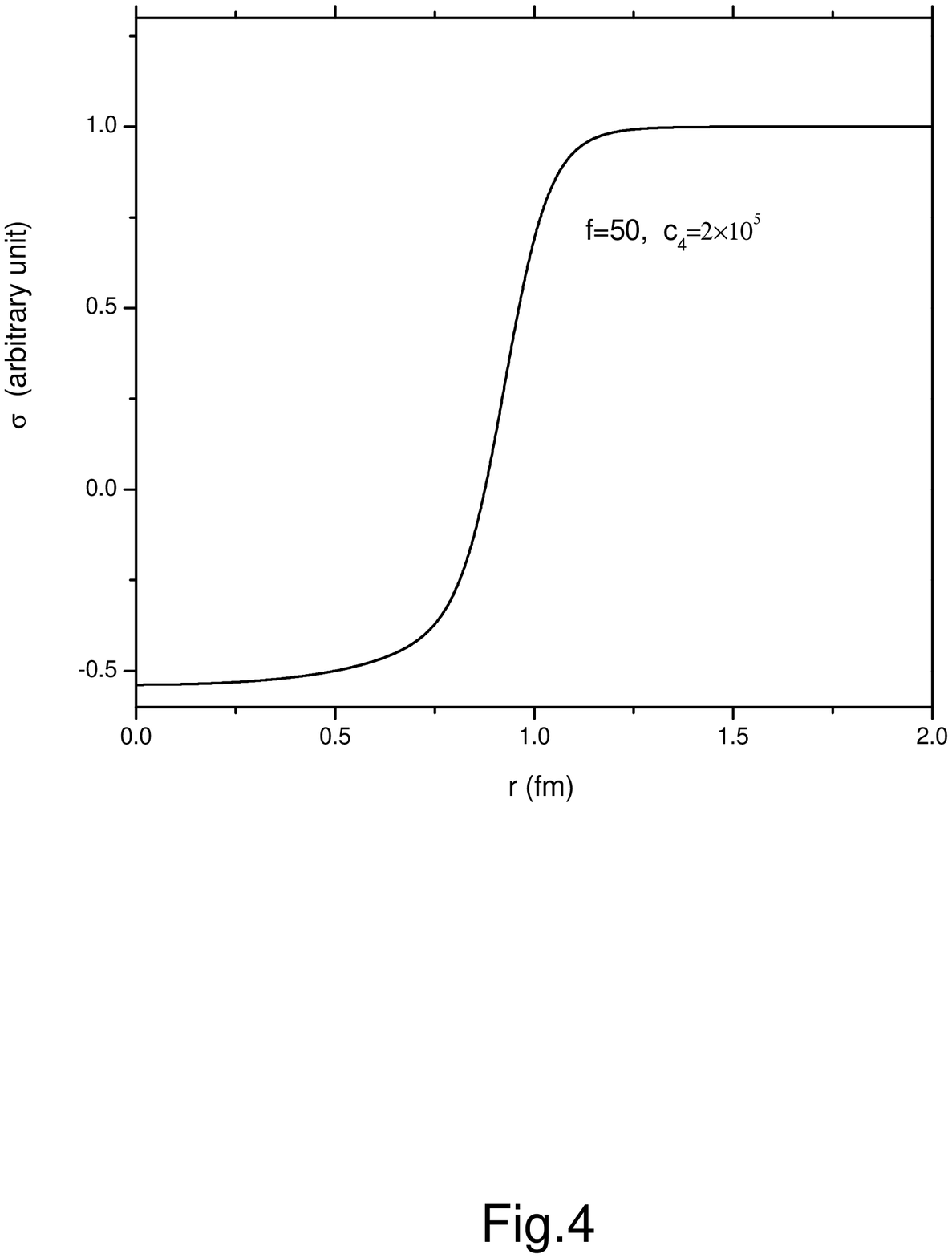}
\caption{The soliton field $\sigma$ versus radius for $f$=50 and
$c_4$=$2\times 10^5$, $B^{1/4}$=145 MeV}
 \label{fig4}
\end{figure}

 \begin{figure}[tbp]
\includegraphics[width=17cm,height=22cm]{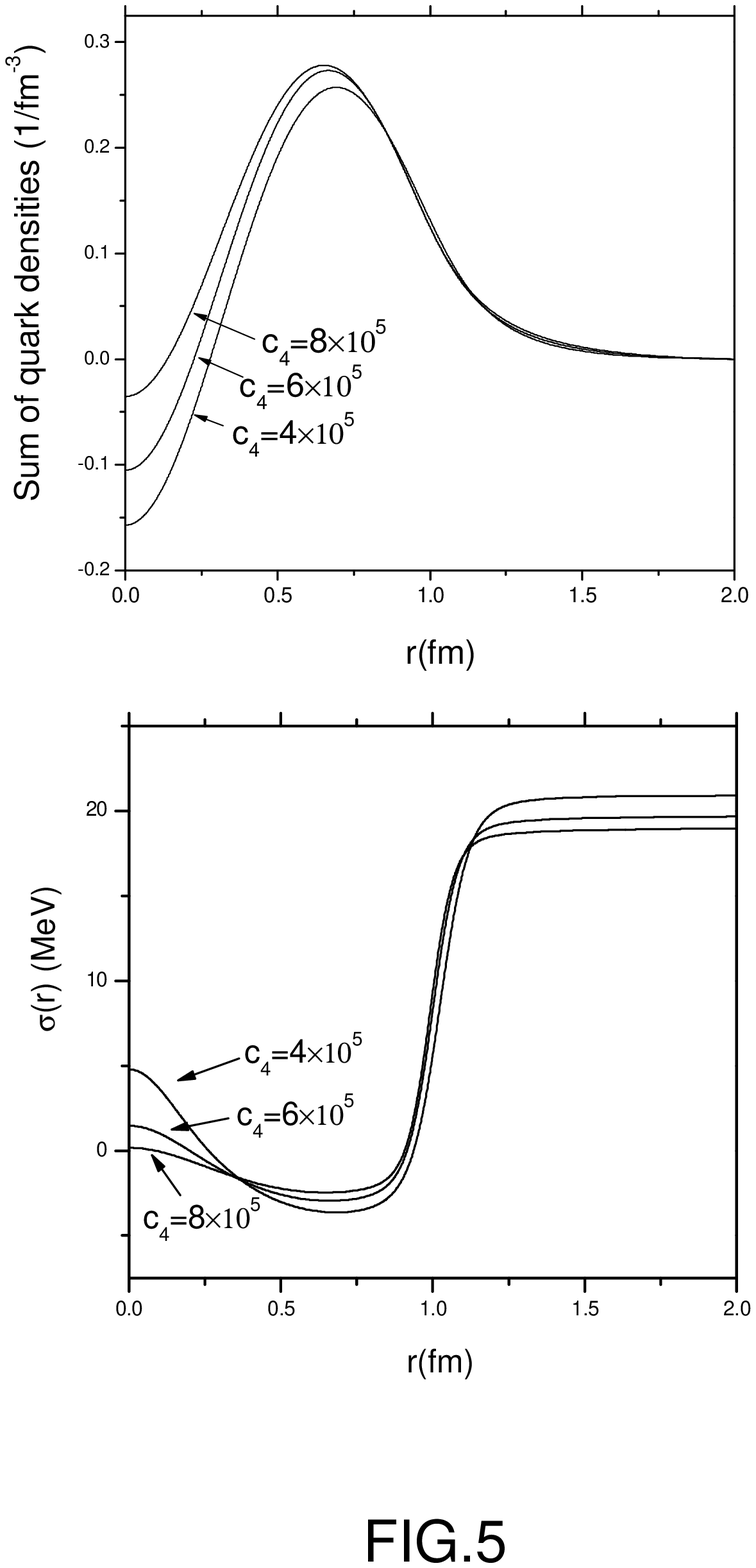}
\caption{Dependence of the odd-parity solutions on the parameter
$c_4$. Other parameters are kept at $f$=30, $r_p$=0.83 fm
 for $B^{1/4}=145$ MeV.
  No solutions exist for smaller values of $c_4$.}
\end{figure}

\begin{figure}[tbp]
\includegraphics[width=17cm,height=22cm]{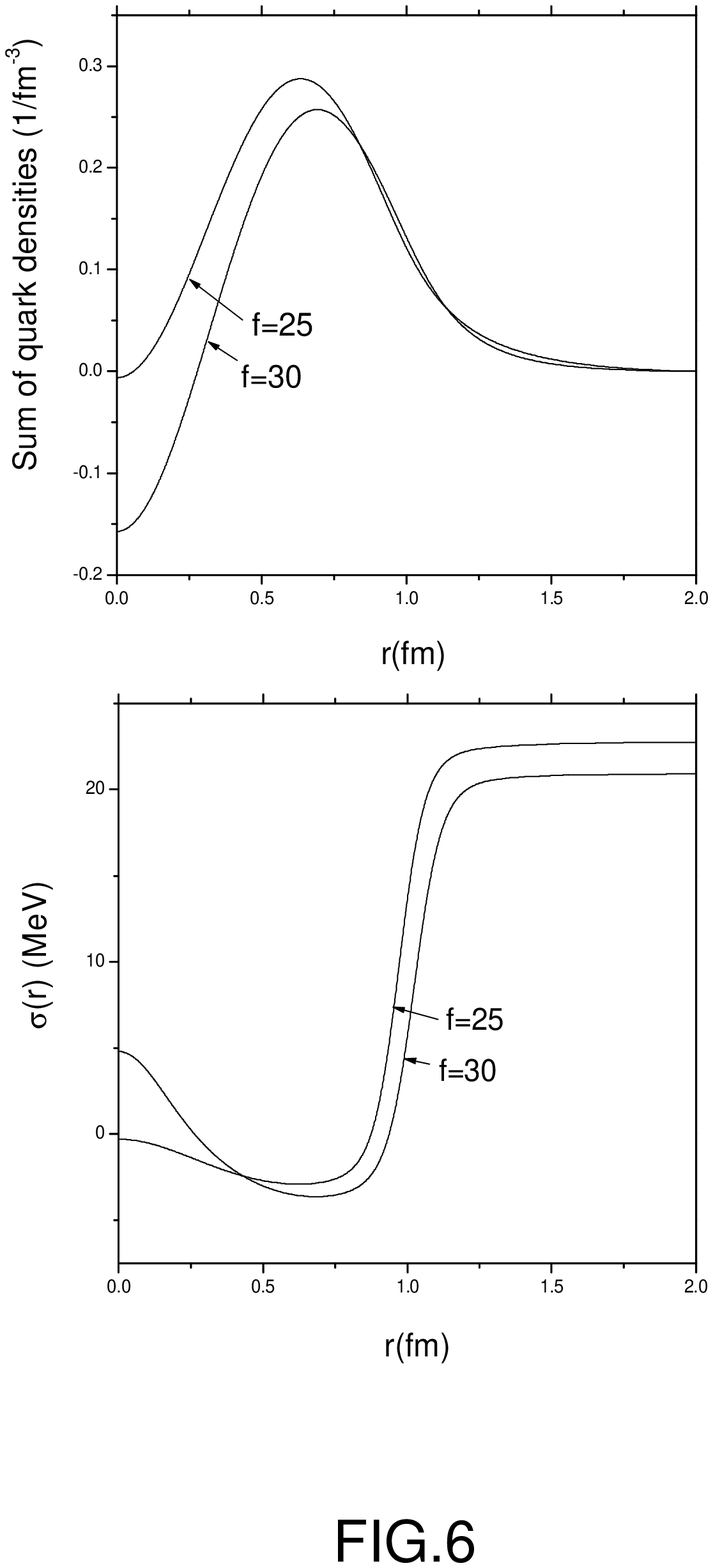}
\caption{Dependence of the odd-parity solutions on the parameter
$c_4$. Other parameters are kept at $f$=30, $r_p$=0.83 fm
 for $B^{1/4}=145$ MeV. No solutions exist for higher values of $f$.}
\end{figure}


\begin{references}
\bibitem{Witten}E. Witten, Phys. Rev. D \textbf{30}, 272 (1984).
\bibitem{Farhi} E. Farhi and R. L. Jaffe, Phys. Rev. D \textbf{30}, 2379 (1984);
               M. S. Berger and R. L. Jaffe, Phys. Rev. C \textbf{35}, 213 (1987);
               E. P. Gilson and R. L. Jaffe, Phys. Rev. Lett \textbf{71}, 332 (1993).
\bibitem{J. madsen}  J. Madsen, Phys. Rev. Lett \textbf{61}, 2909 (1993); Phys. Rev. D \textbf{47}, 5156 (1993); \textbf{50}, 3328 (1994).
\bibitem{B.C.Parija}B. C. Parija, Phys. Rev. C \textbf{53}, 2483 (1993); \textbf{51}, 1473 (1995).
\bibitem{P. Wang}P. Wang, R. K. Su, H. Q. Song, and L. L. Zhang, Nucl. Phys. A \textbf{653}, 166 (1999);
H. Q. Song, R. K. Su, D. H. Lu and W. L, Qian, Phys. Rev. C
\textbf{68}, 055201 (2003).

\bibitem{Y. J. Zhang}Y. J. Zhang, S. Gao, R. K. Su and X. Q. Li,
Chin. Phys. Lett \textbf{14}, 89 (1997).
\bibitem{P.Wang}P. Wang, Z. Y. Zhang, Y. W. Yu, R. K. Su, and H.
Q. Song, Nucl. Phys. A \textbf{688}, 791 (2001).

\bibitem{G. N. Fowler}G. N. Fowler, S. Raha, and R. M. Weiner, Z. Phys. C \textbf{9}, 271 (1981).
\bibitem{Peng}G. X. Peng, H. C. Chiang, B. S. Zou, P. Z. Ning and S. J. Luo, Phys. Rev. C \textbf{62}, 025801 (2000).
\bibitem{S.C}S. Chakrabarty, Phys. Rev. D \textbf{43}, 627 (1991);
ibid \textbf{48}, 1409 (1993).
\bibitem{Lugones}O. G. Benrenuto and G. Lugones, Phys. Rev. D \textbf{51}, 1989 (1995).
\bibitem{zs} Y. Zhang and R. K. Su, Phys. Rev. C \textbf{65}, 035202 (2002); Phys. Rev. C \textbf{67}, 015202(2003);
Europhys. Lett. \textbf{56}, 361 (2001); J. Phys. G \textbf{30},
811 (2004); Mod. Phys. Lett. A \textbf{18}, 143 (2003).
\bibitem{Gupta}V. K. Gupta et al, Int. J. Mod. Phys. D \textbf{21}, 583 (2003).

\bibitem{PAM}P. A. M. Guichon, Phys. Lett. B \textbf{200}, 235 (1988).
\bibitem{PAM}K. Saito and A. W. Thomas, Phys. Lett. B \textbf{327}, 9 (1994).

\bibitem{X. Jin}X. Jin and B. K. Jennings, Phys. Lett. B \textbf{374}, 1 (1996); Phys. Rev. C \textbf{54}, 1427 (1996).

\bibitem{R.Friedberg}R. Friedberg and T. D. Lee, Phys. Rev. D \textbf{15}, 1694 (1977); \textbf{16}, 1096 (1977); \textbf{16}, 1623 (1978).
\bibitem{R.Goldflam}R. Goldflam and L. Wilets, Phys. Rev. D \textbf{25}, 1951 (1982).
\bibitem{Raly}S. Raly and K. Sundaresan, Phys. Rev. D \textbf{29},
525 (1984).

\end{references}
\end{document}